\documentclass[prd,twocolumn,nofootinbib,letter,superscriptaddress]{revtex4-1}

\usepackage{color}
\usepackage{amsmath}
\usepackage{graphicx}
\usepackage{enumerate}
\usepackage{esint}
\usepackage{slashed}
\usepackage[english]{babel}
\usepackage{hyperref}

\hypersetup{
    pdfnewwindow=true,      
    colorlinks=true,       
    linkcolor=black,          
    citecolor=blue,        
    filecolor=blue,      
    urlcolor=blue           
}

\begin{document}

\title{Constraints on long-lived light scalars with flavor-changing couplings \\ and the KOTO anomaly}


\author{P. S. Bhupal Dev}
\email{bdev@wustl.edu}
\affiliation{Department of Physics and McDonnell Center for the Space Sciences,  \\
Washington University, St. Louis, MO 63130, USA}

\author{Rabindra N. Mohapatra}
\email{rmohapat@umd.edu}
\affiliation{Maryland Center for Fundamental Physics, Department of Physics, \\University of Maryland, College Park, MD 20742, USA}

\author{Yongchao Zhang}
\email{yongchao.zhang@physics.wustl.edu}
\affiliation{Department of Physics and McDonnell Center for the Space Sciences,  \\
Washington University, St. Louis, MO 63130, USA}
\affiliation{Department of Physics, Oklahoma State University, \\
Stillwater, OK  74078, USA}


\begin{abstract}
  Recently, the KOTO experiment at J-PARC has observed three anomalous events in the flavor-changing rare decay $K_L \to \pi^0 \nu \bar\nu$, which indicates that the corresponding branching ratio is almost two orders of magnitude larger than the Standard Model (SM) prediction. Taking this intriguing result at face value, we explore model implications of its viable explanation by a long-lived light SM-singlet scalar ($S$) emission, i.e.~$K_L \to \pi^0 S$, with $S$ decaying outside the KOTO detector. We derive constraints on the parameter space of such a light scalar in the context of three simple models: (i) a real singlet scalar extension of the SM; (ii) a $B-L$ extension where neutrino masses arise via type-I seesaw mechanism from $B-L$ breaking;  and (iii) a TeV-scale left-right symmetric model. The flavor-changing couplings needed to explain the KOTO excess in models (i) and (ii) originate from tree-level mixing of the scalar with SM Higgs field ($h$), and in model (iii), from the mixing of $S$ and $h$ with the neutral component of the heavy bidoublet Higgs field. After taking into account the stringent constraints from high-precision searches for flavor-changing charged and neutral kaon decays at NA62, E949, KOTO and CHARM experiments, as well as the astrophysical and cosmological constraints on a light scalar, such as those from supernova energy loss, big bang nucleosynthesis and relativistic degrees of freedom, we find that the light scalar interpretation of the KOTO excess is allowed in all these models. Parts of the parameter range can be tested in future NA62 and DUNE experiments.
\end{abstract}


\maketitle

\section{Introduction}\label{sec:intro}

In the Standard Model (SM) of particle physics, flavor-changing neutral currents (FCNCs) are absent at tree-level and are predicted to be small at loop level, suppressed by the Glashow-Iliopoulos-Maiani (GIM) mechanism  and the small off-diagonal Cabibbo-Kobayashi-Maskawa (CKM) matrix elements in the quark sector (or by the tiny neutrino masses in the lepton sector, if we allow nonzero neutrino masses to be part of the `new' SM)~\cite{PDG}. Any observation of FCNC above the SM prediction would therefore be a clear signature of beyond the SM (BSM) physics. Very recently, the KOTO experiment at J-PARC has observed four candidate events in the signal region of one such rare flavor-changing decay $K_L \to \pi^0 \nu \bar\nu$~\cite{anomaly}. While one of the events is suspected to have originated from SM activity upstream from the detector and can be vetoed away, the remaining three events cannot be explained by currently known backgrounds, with the SM expectation of only $0.05\pm 0.02$ events. This corresponds to a decay branching ratio (BR) of~\cite{anomaly}
\begin{eqnarray}
\label{eqn:anomaly}
{\rm BR} (K_L \to \pi^0 \nu \bar\nu)_{\rm KOTO19} \ = \
2.1_{-1.1(-1.7)}^{+2.0(+4.1)} \times 10^{-9} \,,
\end{eqnarray}
at 68\:(95)\% confidence level (CL), where the uncertainties are primarily due to statistics. This result is consistent with their previously reported 90\% CL upper bound of~\cite{Ahn:2018mvc}
\begin{eqnarray}
\label{eqn:KL1}
{\rm BR} (K_L \to \pi^0 \nu \bar\nu)_{\rm KOTO18} \ < \ 3.0\times 10^{-9} \, .
\end{eqnarray}
The central value in Eq.~\eqref{eqn:anomaly} is almost two orders of magnitude larger than the SM prediction of~\cite{Buras:2015qea}
\begin{eqnarray}
{\rm BR} (K_L \to \pi^0 \nu \bar\nu)_{\rm SM} \ = \
\big( 3.4 \pm 0.6 \big) \times 10^{-11} \, .
\end{eqnarray}
Needless to say, more experimental information on the source of these intriguing events, as well as a careful reevaluation of background estimations, is needed to confirm whether the signal is indeed due to some BSM physics. But given the far-reaching consequences, we take the KOTO result~\eqref{eqn:anomaly} at face value and explore possible implications for some simple BSM scenarios that can be independently tested in other ongoing or future experiments.

At the phenomenological level, the KOTO signal can be interpreted as the emission  of a new light, long-lived scalar particle $S$ in the two-body kaon decay $K_L\to \pi^0S$, which subsequently decays outside the KOTO detector, thus mimicking the invisible $\nu\bar{\nu}$ final states in $K_L\to \pi^0\nu\bar{\nu}$~\cite{Fuyuto:2014cya, Kitahara:2019lws, Egana-Ugrinovic:2019wzj}.\footnote{Other interpretations in terms of either a heavy mediator or a new light particle produced at fixed target and decaying off-axis to two photons (e.g. an axion-like particle) have also been discussed~\cite{Kitahara:2019lws}. Similarly, Ref.~\cite{Fabbrichesi:2019bmo} has considered the possibility of $K_L\to \pi^0Q\bar{Q}$, where $Q$ is a dark fermion of the dark sector. }  In this paper, we consider possible ultraviolet (UV)-complete model frameworks for such a new light scalar particle $S$ with $m_S< m_K-m_{\pi^0}$ and with a flavor-changing effective coupling of the form $K\pi S$ so that it can be emitted in kaon decay.
Being light, there are stringent  constraints on this particle from laboratory searches for FCNCs in the $K$, $D$ and $B$ meson decays. In particular, the scalar emission through the effective $K\pi S$ coupling contributes to both neutral and charged kaon decays, i.e. $K_L\to \pi^0S$ and $K^+\to \pi^+S$, whose branching ratios (for an invisible $S$) are correlated by the Grossman-Nir bound~\cite{Grossman:1997sk}
\begin{eqnarray}
\label{eqn:GN}
{\rm BR} (K_L \to \pi^0 \nu \bar\nu) \ \leq \ 4.3 \,
{\rm BR} (K^+ \to \pi^+ \nu \bar\nu) \,.
\end{eqnarray}
No excess has been reported in the charged kaon decay mode $K^+ \to \pi^+ \nu \bar\nu$, whose branching ratio is currently constrained by the NA62 experiment~\cite{NA62} (and also by E949 experiment~\cite{Artamonov:2009sz}) to be
\begin{eqnarray}
\label{eqn:NA62}
{\rm BR} (K^+ \to \pi^+ \nu \bar\nu)_{\rm NA62} \ < \
2.44 \times 10^{-10} \,
\end{eqnarray}
at 95\% CL, which is consistent with the SM prediction of~\cite{Buras:2015qea}
\begin{eqnarray}
{\rm BR} (K^+ \to \pi^+ \nu \bar\nu)_{\rm SM} \ = \
(8.4\pm 1.0)\times 10^{-11} \ .
\end{eqnarray}
There also exist stringent constraints on a light, long-lived scalar decaying into charged lepton or photon pairs from the searches for $\ell^+\ell^-$ and $\gamma\gamma$ in rare kaon decays at proton beam-dump experiments, such as CHARM~\cite{Bergsma:1985qz}. In addition,  a light $S$ particle will be constrained by astrophysical and cosmological observations, such as those from supernova energy loss, and effective relativistic degrees of freedom ($\Delta N_{\rm eff}$) and/or additional energy injection at the big bang nucleosynthesis (BBN) epoch.

To see whether the KOTO excess is consistent with all these constraints, it is convenient to work within specific models so that the new scalar interactions with SM particles have a definite profile. Due to the suppressed nature of these interactions, only models where the particle $S$ has no direct tree-level coupling to SM quarks need to be considered. We find the following three BSM scenarios which fall into this category:
\begin{enumerate}[(i)]
  \item {\bf Scalar singlet model:} Here the FCNC couplings of $S$ arise from its mixing with the SM Higgs field $h$, which has loop-induced FCNC couplings with SM quarks ~\cite{Willey:1982ti, Grinstein:1988yu, Chivukula:1988gp}. The new scalar $S$ could be long-lived if the $S-h$ mixing angle, $\theta$ is suitably small, while avoiding all existing laboratory constraints~\cite{OConnell:2006rsp, Batell:2009jf, Filimonova:2019tuy, Clarke:2013aya, Winkler:2018qyg, Boiarska:2019jym}. This is a simple, two-parameter model with only $m_S$ and $\theta$ as the unknown parameters. We call this the SM+$S$ model.

  \item {\bf $U(1)_{B-L}$ model:} A class of UV-complete models where such a light scalar without tree-level coupling to SM fermions emerges naturally is based on the gauge group $SU(2)_L \times U(1)_{I_{3R}} \times U(1)_{B-L}$~\cite{Davidson:1978pm, Marshak:1979fm, Mohapatra:1980qe, Deshpande:1979df}. In this case, three right-handed neutrinos (RHNs) $N_a({\bf 1},1/2,-1)$ (with $a=1,2,3$) are introduced for the purpose of anomaly cancellation. The light scalar $S$ can be identified as the real part of a complex $(B-L)$-charged scalar $\Delta({\bf 1},-1,2)$ that breaks the $U(1)_{I_{3R}}\times U(1)_{B-L}$ symmetry to $U(1)_Y$ of the SM and gives mass to the RHNs to implement the type-I seesaw mechanism~\cite{seesaw1,seesaw2,seesaw3,seesaw4,seesaw5}. The long-lived property and the FCNC constraints of this model on $S$ have already been studied in great detail in Ref.~\cite{Dev:2017dui} (where $S$ was denoted by $H_3$), which will be relied upon here. This model has some new, suppressed decay modes such as $S\to NN, Z'Z'$ (where $Z'$ is the massive gauge boson associated with the $U(1)_{B-L}$ breaking) which are absent in the SM+$S$ model. However, as long as all the three RHNs and the $Z'$ boson are much heavier than the light scalar $S$, they will not have any effect on the lifetime of $S$, and the KOTO phenomenology of light $S$ in the $U(1)_{B-L}$ extension will be the same as in the SM+$S$ model. In what follows, we assume this to be the case and therefore do not separately discuss the $U(1)_{B-L}$ scenario for the KOTO explanation, except for the complementary collider signatures, which are different in the $U(1)_{B-L}$ case due to the additional gauge-portal production.

  \item {\bf Left-right symmetric model:} The last class of models studied here is  the left-right symmetric model (LRSM) based on the gauge group $SU(2)_L \times SU(2)_R \times U(1)_{B-L}$~\cite{LR1,LR2,LR3}. Here the light scalar $S$ (denoted by $H_3$ in Refs.~\cite{Dev:2016vle, Dev:2017dui}) can be identified as the real part of the  neutral component of the $(B-L)$-charged, $SU(2)_R$-triplet field $\Delta_R({\bf 1},{\bf 3},2)$, which can be light and does not couple directly to SM quark fields prior to symmetry breaking~\cite{Dev:2016vle, Zhang:2007da, Dev:2016dja}. It is therefore similar to the SM+$S$ model in many respects and can play a role in resolving the KOTO anomaly. The field $\Delta_R$ is responsible for the $SU(2)_R\times U(1)_{B-L}$ symmetry breaking and the model, like the $U(1)_{B-L}$ model above, has the extra motivation of being connected to neutrino mass generation via type-I seesaw~\cite{seesaw1,seesaw2,seesaw3,seesaw4,seesaw5}. 
 In contrast with the previous two models, the FCNC couplings of $S$ in this case arise at tree-level, due to its mixing with the heavy scalar $H_1$ from the bidoublet $\Phi$ (and the SM Higgs). Another special feature of the light scalar $S$ in the LRSM is that even for small mixing angles , it can still decay into two photons through the $W_R$ loop and the heavy charged scalar loops. This makes the FCNC limits, as well as the supernova and BBN limits, on light $S$ in the LRSM very different from the other two models discussed above.
\end{enumerate}
As we show below, a limited parameter space that satisfies all the existing constraint  allows an explanation of the KOTO excess in  all the models.
This allowed parameter range can be tested in the future high-precision intensity frontier experiments, such as NA62 and DUNE.

The rest of the paper is organized as follows: In Sec.~\ref{sec:generic}, we discuss the simplest real scalar extension of the SM in light of the KOTO excess vis-\'{a}-vis other laboratory and astrophysical/cosmological constraints. Most of this discussion is also applicable to the $U(1)_{B-L}$ case. In Sec.~\ref{sec:lrsm}, we repeat the same exercise for the LRSM. Our conclusions are given in Sec.~\ref{sec:con}.

\section{Singlet model}
\label{sec:generic}


The singlet scalar extension of the SM is one of the simplest and well-motivated BSM scenarios~\cite{OConnell:2006rsp}.
The most general renormalizable scalar potential of the SM Higgs doublet $H$ and a real singlet scalar $S$ can be written as
\begin{eqnarray}
\label{eqn:potential}
V_{} & \ = \ &
-\mu_1^2 (H^\dagger H) - \mu_2^2 S^2 \nonumber \\
&& + \lambda_1 (H^\dagger H)^2
+ \lambda_2 H^\dagger H S^2
+ \lambda_3 S^4 \,,
\end{eqnarray}
with $\mu_{1,2}^2 >0$ being the mass parameters and $\lambda_{1,2,3}$ being the quartic couplings. We impose a $Z_2$-symmetry under which $S\to -S$ to prevent the $S^3$  and $SH^\dagger H$ trilinear terms. After spontaneous symmetry breaking, the $H$ and $S$ fields obtain non-vanishing vacuum expectation values (VEVs), with $\langle H \rangle = (0, v_{\rm EW})^{\sf T}$ with $v_{\rm EW} \simeq 174$ GeV, the electroweak (EW) VEV and $\langle S \rangle = v_S$. The $h - S$ mixing (where $h$ is the physical SM Higgs field, obtained by expanding the $H$-field around its VEV, i.e. $H= (0, v_{\rm EW}+h)^{\sf T}$) is determined by the quartic coupling $\lambda_2$. In the small mixing limit, the mass of the real component of $S $ is $m_{S}^2 \simeq 4\lambda_3 v_S^2$ to the leading order. For sufficiently small $\lambda_3$ and $v_S$, the scalar $S$ could be very light, even down to a few MeV scale.\footnote{When $S$ is light, the SM Higgs might contribute radiatively to the $S$ mass, potentially making it heavier. However, this effect is highly suppressed by the $h-S$ mixing angle $\sin\theta$, which needs to be small to make the $S$ long-lived, as required for the KOTO excess explanation.}

In the $U(1)_{I_{3R}}\times U(1)_{B-L}$ extension discussed in Sec.~\ref{sec:intro}, the $S$-field can be identified as the real part of a $(B-L)$-charged scalar, whose VEV breaks the $U(1)_{I_{3R}}\times U(1)_{B-L}$ gauge symmetry down to $U(1)_Y$~\cite{Dev:2017dui}. We assume the RHNs in this case are all heavier than $S$, so that the decays of $S$ are identical to those of the SM+$S$ case, being governed only by two parameters, namely, the scalar mass $m_S$ and the $h-S$ mixing angle $\sin\theta$.



\subsection{Fitting the KOTO anomaly}\label{sec:koto1}

The couplings of $S$ to the SM fermions arise from its mixing with the SM Higgs $h$ and are thus flavor-conserving at the tree level. However, FCNCs are generated at one-loop level through Penguin diagrams involving the $W-$top loop and CKM quark mixings. The effective Lagrangian relevant for FCNC kaon decay is given by~\cite{Batell:2009jf}
\begin{eqnarray}
&& {\cal L}_{\rm eff} \ \supset \ y_{sd} \sin\theta S \bar{s}_L d_R + {\rm H.c.} \,, \quad \nonumber \\
&& \text{with }
y_{sd} \ = \ \frac{3\sqrt2 G_F m_t^2 V_{ts}^\ast V_{td}}{16\pi^2} \,
\frac{m_S}{\sqrt2 v_{\rm EW}} \, ,
\label{eqn:ysd}
\end{eqnarray}
where $y_{sd}$ is the effective loop-level coupling in the SM, $G_F$ is the Fermi constant,
$m_{s,\,t}$ the strange and top quark masses, and $V_{td,\, ts}$ the CKM matrix elements. As a result of the $CP$ phase in the CKM matrix, the coupling $y_{sd}$ is complex. If kinematically allowed, this will induce the flavor-changing decays $K^\pm \to \pi^\pm S$ and $K_L \to \pi^0 S$, with the partial widths
\begin{eqnarray}
\label{eqn:Kdecay:U1:1}
\Gamma (K^\pm \to \pi^\pm S) &\ \simeq \ &
\frac{m_{K^\pm} \left| y_{sd} \right|^2 \sin^2\theta}{64 \pi}
\frac{m_{K^\pm}^2}{m_S^2} \nonumber \\
&& \times \beta_2 (m_{K^\pm}, m_{\pi^\pm}, m_{S}) \,, \\
\label{eqn:Kdecay:U1:2}
\Gamma (K_L \to \pi^0 S) &\ \simeq \ &
\frac{m_{K_L} \left( {\rm Re} \, y_{sd} \right)^2 \sin^2\theta}{64 \pi}
\frac{m_{K^0}^2}{m_S^2} \nonumber\\
&& \times \beta_2 (m_{K_L}, m_{\pi^0}, m_{S}) \,,
\end{eqnarray}
with 
the kinematic function
\begin{eqnarray}
\label{eqn:beta2}
\beta_2 (M,\, m_1,\, m_2) & \ \equiv \ & \left[ 1 - \frac{2(m_1^2 + m_2^2)}{M^2} + \frac{(m_1^2 - m_2^2)^2}{M^4} \right]^{1/2} \,. \nonumber \\ &&
\end{eqnarray}
Note that the partial decay widths in Eqs.~(\ref{eqn:Kdecay:U1:1}) and (\ref{eqn:Kdecay:U1:2}) are almost identical, except for the crucial difference that the decay $K_L \to \pi^0 S$ depends only on the real part of the coupling $y_{sd}$.

The same $h-S$ mixing is also responsible for $S$ decays into the SM quarks ($u,d,s$) and charged leptons at tree-level, and gluons and photons at one-loop level, just like the SM Higgs boson decays. In the generic singlet model all the decay modes of $S$ are universally proportional to the mixing angle $\sin\theta$, and therefore, the branching ratios depend only on the $S$ mass but not on $\sin\theta$. As detailed in Ref.~\cite{Dev:2017dui}, if $S$ is light, say below the GeV-scale, it tends to be long-lived for a wide range of $\sin\theta$. If its average decay length is larger than the KOTO detector size, the process of interest will be
\begin{eqnarray}
K_L \to \pi^0 S \,, \quad S \to \text{invisible} \, ,
\end{eqnarray}
with $S$ decaying into anything outside the detector. This has the same final state as the decay $K_L \to \pi^0 \nu \bar\nu$, i.e. two photons from $\pi^0 \to \gamma\gamma$ and significant missing energy. In this case, the effective branching ratio is given by\footnote{There is an ${\cal O}(1)$ correction factor to account for the kinematical difference between the 3-body SM decay $K_L\to \pi \nu\bar{\nu}$ and the 2-body decay $K_L\to \pi^0 S$ in our scalar case, whose exact value depends on the scalar mass~\cite{Ahn:2018mvc}. Here we have simply assumed it to be one, given the fact that there is no directional information in the KOTO signal which only involves charge-neutral particles and vetoes all charged particles.}
\begin{eqnarray}
\label{eqn:BReff}
{\rm BR}^{\rm eff} (K_L \to \pi^0 S) \ = \
{\rm BR} (K_L \to \pi^0 S) \exp [ - L \Gamma_{S}/b ] \,, \nonumber \\
\end{eqnarray}
where ${\rm BR} (K_L \to \pi^0 S) = \Gamma (K_L \to \pi^0 S)/\Gamma^{\rm total}_{K_L}$, $L =3$ m for the KOTO detector, and $b= E_{S}/m_{S}$ the Lorentz boost factor with energy $E_{S} \simeq 1.5$ GeV. For the total decay width of $K_L$, we use $\Gamma^{\rm total}_{K_L}=\Gamma (K_L \to \pi^0 S)+\Gamma^{\rm SM}_{K_L}$, where $\Gamma (K_L \to \pi^0 S)$ is given by Eq.~\eqref{eqn:Kdecay:U1:2} and $\Gamma^{\rm SM}_{K_L}=(1.29\pm 0.01)\times 10^{-17}$ GeV~\cite{PDG}.

\begin{figure*}[!t]
  \centering
  \includegraphics[width=0.65\textwidth]{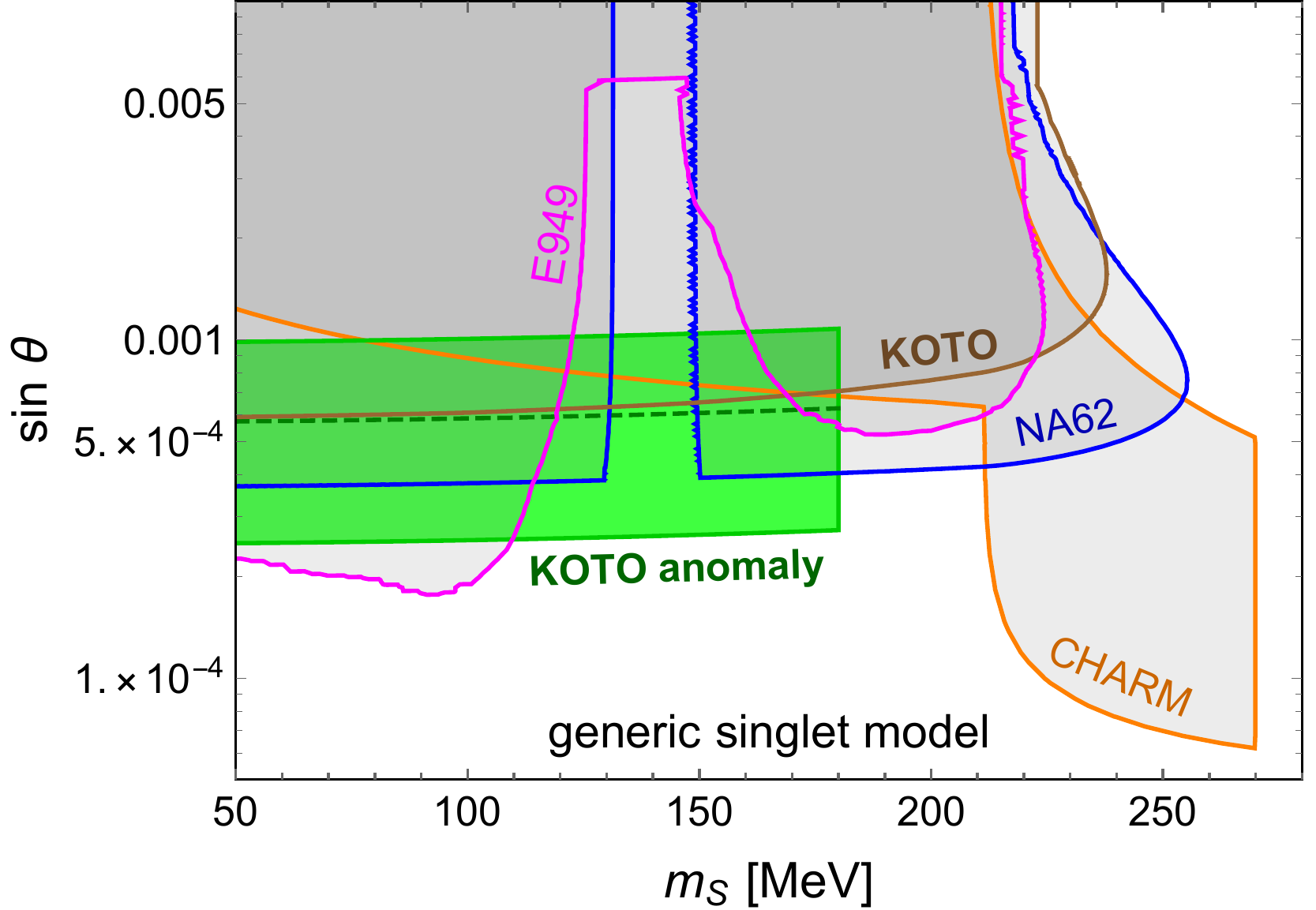}
  \caption{The parameter space favored by the KOTO anomaly~\cite{anomaly} in the generic singlet scalar model (cf.~Sec.~\ref{sec:koto1}) is shown by the green shaded region (95\% CL), with the green dashed line corresponding to the central value quoted in Eq.~\eqref{eqn:anomaly}. For comparison, we show the exclusion regions from a previous KOTO search for $K_L\to \pi^0\nu\bar{\nu}$
  (brown)~\cite{Ahn:2018mvc}, NA62 search for $K^+ \to \pi^+ \nu \bar\nu$ (blue)~\cite{NA62}, E949 search for $K^+ \to \pi^+ X$ (magenta)~\cite{Artamonov:2009sz}, and beam-dump experiment at CHARM  (orange)~\cite{Bergsma:1985qz}; cf.~Sec.~\ref{sec:lab1}. All the gray shaded regions are excluded.
  }
  \label{fig:U1}
\end{figure*}

Using Eq.~\eqref{eqn:BReff}, we calculate the preferred region in the $(m_S,\sin\theta)$ parameter space that explains the KOTO excess given by Eq.~\eqref{eqn:anomaly} at 95\% CL. Our result is shown by the green shaded region in Fig.~\ref{fig:U1}, with the green dashed line corresponding to the KOTO central value in Eq.~\eqref{eqn:anomaly}. The region with $m_S > 180$ MeV is not included in this fit, because it does not overlap with the KOTO signal region~\cite{Kitahara:2019lws}.



\subsection{Laboratory constraints} \label{sec:lab1}

As shown  by the Grossman-Nir bound [cf.~Eq.~\eqref{eqn:GN}], the FCNC decays of charged and neutral kaons are correlated. This relation has to do with general isospin symmetry arguments, which relate the decay amplitudes of $K^\pm$ to those of $K^0$ and $\overline{K^0}$, and holds even for the 2-body decays $K_L\to \pi^0S$ and $K^+\to \pi^+S$~\cite{Leutwyler:1989xj}. For our singlet scalar case, this can be explicitly seen from Eqs.~(\ref{eqn:Kdecay:U1:1}) and (\ref{eqn:Kdecay:U1:2}). As a result, the current and future high-precision measurements of the charged and neutral kaon rare decays can be used to set limits on the scalar mass $m_{S}$ and mixing angle $\sin\theta$, as discussed below.

In the charged kaon sector, the most stringent limits come from searches of $K^+ \to \pi^+ \nu \bar\nu$ at NA62~\cite{NA62} and of $K^+ \to \pi^+ X$ (with $X$ being a long-lived particle) at E949~\cite{Artamonov:2009sz}. The NA62 experiment has put a 95\% CL upper limit on ${\rm BR}(K^+ \to \pi^+ \nu \bar\nu)$ [cf.~Eq.~\eqref{eqn:NA62}] which can be translated into an exclusion region in the $(m_S,\sin\theta)$ plane, as shown by the blue line in Fig.~\ref{fig:U1}. Here we have constructed an effective BR, similar to Eq.~\eqref{eqn:BReff}, replacing neutral mesons by charged mesons and modifying the experimental parameters to $L=150$ m and $E_S\sim 37$ GeV for NA62. Again we have neglected the ${\cal O}(1)$ kinematical difference between the 3-body SM decay and the 2-body decay in our scalar case. We see from Fig.~\ref{fig:U1} that there is a gap in the NA62 excluded region around the pion mass. This is because of the fact that if the scalar mass $m_S$ is close to $\pi^0$ mass, we will have a large pion background from the SM process $K^+ \to \pi^+ \pi^0$ with $\pi^0 \to \nu \bar\nu$. With the current limit of ${\rm BR} (\pi^0 \to \nu \bar\nu) < 2.7 \times 10^{-7}$~\cite{PDG} and the SM $K^+\to \pi^+\pi^0$ branching ratio of 20.6\%~\cite{PDG}, the NA62 limit on $K^+ \to \pi^+S$ turns out to be two orders of magnitude weaker in this region, as shown by the gap in Fig.~\ref{fig:U1}.

The E949 experiment has reported 95\% CL bounds on ${\rm BR}(K^+ \to \pi^+ X)$, where $X$ is a long-lived particle, as a function of the $X$ mass~\cite{Artamonov:2009sz}.  The most stringent limit on the branching ratio ${\rm BR}(K^+ \to \pi^+ X)$ could reach up to $5.4\times 10^{-11}$ if the new particle $X$ is stable. Using the same procedure as above for NA62, we evaluate the effective branching ratio following Eq.~(\ref{eqn:BReff}), with decay length $L = 4$ m and energy $E_{S} \simeq 710$ MeV for E949. The corresponding exclusion region is shown by the magenta line in Fig.~\ref{fig:U1}, which is up to a factor of few stronger than the NA62 exclusion region in the low-mass range, but is weaker in the high-mass range and not applicable for $m_S>2m_{\pi^0}$ because in this case the $S$ tends to decay quickly, compared to the E949 detector size of 4 m. Like the NA62 limit, there is also a gap for the E949 constraint when $m_S \sim m_\pi$.

Similarly, a previous KOTO search has reported 90\% CL upper limits on the 2-body decay ${\rm BR} (K_L \to \pi^0 X)$, where $X$ is an invisible boson, as a function of the $X$ mass~\cite{Ahn:2018mvc}. We can directly use this bound for our scalar case, and following Eq.~\eqref{eqn:BReff} with $L = 3$ m and $E_{S} \simeq 1.5$ GeV for KOTO, translate it into an exclusion region in the $(m_S,\sin\theta)$ plane, as shown in Fig.~\ref{fig:U1} by the brown line. Note that there is no gap in the KOTO limit for $m_S\sim m_\pi$, because the 2-body decay of $K_L^0\to \pi^0\pi^0$ is $CP$-violating and CKM-suppressed in the SM, with a branching ratio of $8.6\times 10^{-4}$~\cite{PDG}.


Further limits on the light scalar can be derived from the $e^+ e^-$, $\mu^+ \mu^-$ and $\gamma\gamma$ decay products of $S$ produced in neutral and charged kaon decays at proton beam-dump experiment such as CHARM~\cite{Bergsma:1985qz}. The production cross section of $S$ at CHARM is given by~\cite{Bezrukov:2009yw,Clarke:2013aya,Dolan:2014ska}
\begin{eqnarray}
\label{eqn:xs}
\sigma_{S} & \ \simeq \ & \sigma_{pp} M_{pp} \left[
\frac12 \chi_s D_{K^\pm} {\rm BR} (K^+ \to \pi^+ S) \right. \nonumber \\
&& \left. + \frac14 \chi_s D_{K_L} {\rm BR} (K^0 \to \pi^0 S)\right] \,,
\end{eqnarray}
where $\sigma_{pp}$ is the proton-proton cross section, $M_{pp}=11$ the average hadron multiplicity, and $\chi_s = 1/7$ is the fraction of strange pair-production rate. In Eq.~\eqref{eqn:xs}, the factor $D_{K} \simeq \ell_K / b_K c \tau_K$ (with $K$ standing for both $K^\pm$ and $K_L$) takes into account the re-absorption of Kaons before decaying~\cite{Egana-Ugrinovic:2019wzj}, with $\ell_K = 15.3$ cm the absorption length, $b_K = E_K/m_K$ the Lorentz boost factor with $E_K \simeq 25$ GeV, and $\tau_K$ the total Kaon decay width. It turns out that the re-absorption factors are respectively $8.1 \times 10^{-4}$ and $2.0 \times 10^{-4}$ for $K^\pm$ and $K_L$.  Normalized to the neutral pion yield $\sigma_{\pi^0} \simeq \sigma_{pp} M_{pp} /3$, we can predict the total number of $S$ particles produced: $N_{S} \simeq 2.9 \times 10^{17} \sigma_{S} / \sigma_{\pi^0}$. Then the number of events collected by the detector would be
\begin{align}
N_{\rm event} \ = \ & N_{S}
\left( \sum_{\chi = e,\mu,\gamma} {\rm BR} (S \to \chi\chi) \right) \nonumber \\
& \times \left[
\exp\left( -\frac{L \Gamma_{S}}{b} \right) -
\exp\left( -\frac{(L+\Delta L) \Gamma_{S}}{b} \right)
\right] \,, 
\end{align}
where $L = 480$ m is the CHARM beam dump baseline, $\Delta L = 35$ m is the detector fiducial length, and $b = E_{S} / m_{S}$ is the boost factor with $E_{S} \simeq E_K/2$~\cite{Bergsma:1985qz}. Due to the huge number of events $N_{S}$, the mixing angle $\sin\theta$ is expected to be severely constrained, and the most stringent limits are from the $ee$ and $\mu\mu$ channels, as the $\gamma\gamma$ channel is comparatively suppressed by the loop factor. Given that no signal event was found at CHARM, an upper limit of $N_{\rm event} < 2.3$ at the 90\% CL on the contribution from BSM physics was set. We use this limit to derive the corresponding exclusion region in the $(m_S,\sin\theta)$ plane, as shown by the orange line in Fig.~\ref{fig:U1}. For lighter $S$, the boost factor $b$ becomes larger, and fewer $S$ decays happen inside the detector, thereby weakening the constraints. As can be seen from Fig.~\ref{fig:U1}, even if all the laboratory constraints are taken into consideration, {there is still a narrow parameter space in the singlet model, i.e. $110 \, {\rm MeV} \lesssim m_S \lesssim 180$ MeV and $2.5 \times 10^{-4} \lesssim {\rm sin}^\theta \lesssim 6.5 \times 10^{-4}$}.



For the sake of completeness, we list here also other limits from the high-precision quark flavor data~\cite{Dev:2017dui} that are either not applicable or weaker than those shown in Fig.~\ref{fig:U1}.
\begin{itemize}
  \item The lifetimes of $K^\pm$ and $K^0$ are both precisely measured up to the level of $10^{-3}$, although the absolute theoretical values are subject to a large uncertainty of strange quark mass, up to the order of 10\%~\cite{PDG}. Therefore, for sufficiently large mixing angles the contribution of $K \to \pi S$ to the total kaon decay widths will suppress the current uncertainties. However these limits from the total decay widths are comparatively much weaker than those from the rare decays discussed above.

  \item The are also some searches of the rare decays $K^+ \to \pi^+ \chi\chi$ with $\chi\chi = e^+ e^- ,\, \mu^+ \mu^-,\, \gamma\gamma$, which have been performed by  NA48/2~\cite{Batley:2009aa, Batley:2011zz} and NA62~\cite{Ceccucci:2014oza}. The neutral lepton decays $K_L \to \pi^0 \chi\chi$ (with $\chi\chi = e^+ e^- ,\, \mu^+ \mu^-,\, \gamma\gamma$) are also searched for in the KTeV experiment~\cite{AlaviHarati:2003mr, AlaviHarati:2000hs, Abouzaid:2008xm}. In these searches, the electrons, muons and photons are all from the primary kaon decay vertex, thus they are not applicable for the long-lived $S$ discussed here for the KOTO anomaly.
  \item The limits from the beam-dump experiment NuCal apply only for a light scalar with mass $m_S \lesssim 80$ MeV~\cite{Blumlein:1990ay}, and they are not relevant to the KOTO anomaly here.
  \item In both the singlet scalar and $U(1)_{B-L}$ models, the loop-level FCNC structure is fixed by the SM quark masses and the CKM matrix [see Eq.~(\ref{eqn:ysd})], thus we can also use the flavor-changing data in the $B$ meson sector to set limits on the $S$ mass and mixing angle $\sin\theta$. However, as the $B$ meson is much heavier than the $K$ meson, the production rate of $B$ mesons is much smaller, and as a result the limits from the rare decays of $B \to K \nu \bar\nu$ at BaBar and Belle are comparatively much weaker than those from the $K$ meson decays, being respectively $3.2 \times 10^{-5}$~\cite{Lees:2013kla} and $1.6 \times 10^{-5}$~\cite{Grygier:2017tzo} (see Fig. 19 of Ref.~\cite{Dev:2017dui}). The future prospects at Belle II~\cite{Abe:2010gxa} are also not comparable to those from the Kaon sector. Furthermore, the searches in the visible channels $B \to K \ell^+ \ell^-$ ($\ell = e,\, \mu$) at BaBar~\cite{Aubert:2003cm}, Belle~\cite{Wei:2009zv} and LHCb~\cite{Aaij:2012vr} are not applicable to the long-lived $S$ case here.

  \item As detailed in Refs.~\cite{Dev:2016vle, Dev:2017dui}, there are also some limits from the measurements of neutral $K$ and $B$ meson oscillations~\cite{PDG}, from ${\rm BR} (B_s \to \mu^+ \mu^-)$ by LHCb~\cite{CMS:2014xfa}, ${\rm BR} (B_d \to \gamma\gamma)$ by BaBar~\cite{delAmoSanchez:2010bx} and ${\rm BR} (B_s \to \gamma\gamma)$ by Belle~\cite{Dutta:2014sxo}. However, these limits are much weaker and are not relevant to the KOTO anomaly.

\end{itemize}

If the light scalar $S$ is long-lived, it can also be searched for at the LHC~\cite{Alimena:2019zri} and/or the dedicated long-lived particle (LLP) detectors such as MATHUSLA~\cite{Curtin:2018mvb}. At the high-energy colliders, $S$ can be produced from the loop-level gluon fusion process $gg \to gS$ via mixing with the SM Higgs, and the cross section can go up to (25 pb) $\times \sin^2 \theta$~\cite{Dev:2017dui}. The LLP searches at the HL-LHC and FCC-hh could probe a large parameter space of $m_S$ and $\sin\theta$ (see Fig.~20 in Ref.~\cite{Dev:2017dui}); however, they do not cover the KOTO parameter space of interest in Fig.~\ref{fig:U1}, and hence, are not shown.

\subsection{Astrophysical and cosmological constraints}
\label{sec:ac1}


For completeness, we consider in this subsection the astrophysical and cosmological constraints on the light long-lived scalar $S$.
A sufficiently light $S$ can be produced in significant amounts in the supernova core via the nuclear bremsstrahlung process ${\cal N} + {\cal N} \ \to \ {\cal N} + {\cal N} + S$,
with ${\cal N} = p,\, n$ collectively standing for protons and neutrons. This process is induced by  the mixing  of $S$ with the SM Higgs field and the effective couplings of the SM Higgs to nucleons. Through the couplings to quarks inside nucleons, the effective couplings $g_{hNN}$ of the SM Higgs to nucleons are of order $\sim 10^{-3}$~\cite{Shifman:1978zn, Cheng:1988im}.  Let us first make a ballpark estimate of the supernova limits. The total energy loss rate due to the emission of the light scalar $S$ is
\begin{eqnarray}
{\cal Q} & \ \sim \ & V_c n_N^2 \sigma_{NN \to NNS} \langle E_S \rangle \nonumber \\
& \ \sim \ & \frac{3V_cn_N^2 \alpha_\pi^2 g_{hNN}^2 T^{7/2} \sin^2\theta}{4 \pi^{3/2} m_N^{9/2}} \,,
\end{eqnarray}
where $V_c = \frac{4\pi}{3} R_c^3$ is the supernova core volume with $R_c$ the core size, $n_N$ the nuclear density in the supernova core, $\alpha_\pi \simeq (2 m_N / m_\pi)^2/4\pi$ the effective coupling of pion to nucleons with $m_{N}$ and $m_\pi$ respectively the masses of nucleons and pion, $T \simeq 30$ MeV the temperature in the supernova core, $\langle E_S \rangle$ the averaged energy of the scalar $S$, and $\sigma_{NN \to NNS}$ the production cross section.
Taking a typical supernova core size $R_c = 10$ km, $n_N = 1.2 \times 10^{38}{\rm cm}^{-3}$,
we get ${\cal Q} \simeq 6 \times 10^{65} \sin^2\theta$ erg/sec. Comparing with the observed energy loss rate of $3\times 10^{53}$ erg/sec~\cite{Hirata:1987hu}, we get $\sin\theta \lesssim 10^{-6}$. However, when the mixing angle $\sin\theta$ between $S$ and the SM Higgs is too large,  the decay lifetime $\tau_S$ of $S$ will be highly suppressed by the mixing angle via $\tau_S \propto \sin^{-2}\theta$ such that the scalar $S$  will decay inside the supernova core and will not contribute to supernova energy loss. Moreover, for the parameter space of $m_S$ and $\sin\theta$ relevant for the KOTO anomaly, the reabsorption of $S$ via the $3\to2$ process ${\cal N} +{\cal N} + S \to {\cal N} + {\cal N}$ turns out to be more important than the decay, as the mean free path (MFP) $\lambda$ is much smaller than the  core size $R_c$.
The calculation of estimating the MFP via the $3\to 2$ reabsorption involves techniques used for the axion~\cite{Burrows:1990pk, Giannotti:2005tn, Krnjaic:2015mbs} (except that we have a more massive scalar particle), and the result is~\cite{DMZ}:
\begin{eqnarray}
\lambda^{-1} & \ = \ & \frac{1}{2E_S} \frac{{\rm d} {\cal B}_S}{{\rm d} \Pi_S} 
\ \sim \ \frac{6 \pi^{1/2} n_N^2 \alpha_\pi^2 g_{hNN}^2 \sin^2\theta}{m_N^{9/2} T^{1/2}} \,,
\end{eqnarray}
where ${\cal B}_S$ is the total number of $S$ produced per unit volume and $\Pi_S$ is the phase space of $S$. Our preliminary results show that for the core temperature of $T \simeq 30$ MeV and a scalar mass $m_S \lesssim 100$ MeV, the MFP is
\begin{eqnarray}
\label{eqn:lambda}
\lambda \ \sim \ 10 \, {\rm km} \times \left( \frac{\sin\theta}{10^{-6}} \right)^{-2} \,.
\end{eqnarray}
The KOTO anomaly favors a mixing angle in the range of $\sin\theta \sim 10^{-4}$--$10^{-3}$. Eq.~(\ref{eqn:lambda}) implies that for such values of mixing angle the MFP is much smaller than the core size of 10 km, and the light scalar $S$ can not take away energy from the supernova core. Therefore, the supernova limits are not applicable to the KOTO-favored region. More details of the supernova limits can be found in Ref.~\cite{DMZ}.


There also exist constraints on such a light $S$ from BBN since it can contribute an extra degree of freedom at that epoch of the universe and spoil the success of BBN. If the mixing angle $\sin\theta$ is very small, say $\lesssim 10^{-5}$ {for a scalar mass of 100 MeV}, the lifetime of a light $S$ might be longer than one second, which would potentially affect BBN in the early universe.
However such small mixing angles are not relevant to the KOTO anomaly here, thus the BBN limit is not shown in Fig.~\ref{fig:U1}. A detailed analysis of the BBN constraint in the context of the singlet scalar model can be found e.g. in Ref.~\cite{Fradette:2017sdd}.

A light $S$ might also contribute to the relativistic degrees of freedom $N_{\rm eff}$ in the early universe, thus getting constrained by the current precision Planck data~\cite{Aghanim:2018eyx}. However, if the mixing angle $\sin\theta$ is too small, $S$ cannot be kept in equilibrium with the SM photon. In particular, if the mixing angle $\sin\theta \lesssim 0.01$ and the scalar mass $m_S\gg$ 1 MeV, the  decay rate $\Gamma (S \to \gamma\gamma) \exp [ -m_{S}/T_\ast ]$ will be Boltzmann suppressed and is significantly smaller than the Hubble expansion rate $H\simeq 10 T_\ast^2/M_{\rm Pl}$, with $T_\ast \sim T_{\rm BBN} \sim $ MeV and $M_{\rm Pl}$ the Planck mass. As the $\Delta N_{\rm eff}$ limit is very weak and not relevant to the KOTO anomaly, we do not show it in Fig.~\ref{fig:U1}.



\section{Left-right symmetric model}
\label{sec:lrsm}


In the minimal version of the LRSM based on the gauge group $SU(2)_L \times SU(2)_R \times U(1)_{B-L}$, there is one bidoublet $\Phi({\bf 2},{\bf 2},0)$ and one right-handed triplet $\Delta_R({\bf 1},{\bf 3},2)$ in the scalar sector:
\begin{eqnarray}
\Phi & \ = \ & \left(\begin{array}{cc}\phi^0_1 & \phi^+_2\\\phi^-_1 & \phi^0_2\end{array}\right) \, , \nonumber \\
\Delta_R & \ = \ & \left(\begin{array}{cc}\Delta^+_R/\sqrt{2} & \Delta^{++}_R\\\Delta^0_R & -\Delta^+_R/\sqrt{2}\end{array}\right)  \, .
\label{eq:scalar}
\end{eqnarray}
The $SU(2)_R \times U(1)_{B-L}$ symmetry is broken down to the SM $U(1)_{Y}$ gauge group, once the triplet develops a non-vanishing VEV $\langle \Delta^0_R \rangle  =  v_R$.  The bidoublet $\Phi$, with the VEVs $\langle \phi^0_1 \rangle  =  \kappa$ and $\langle \phi^0_2 \rangle =  \kappa'$ (where  $v_{\rm EW} = \sqrt{\kappa^2 + \kappa^{\prime \, 2}}$), is responsible for breaking the SM gauge group $SU(2)_L \times U(1)_Y$ down to $U(1)_{\rm em}$ and for the generation of SM quark and charged lepton masses as well as the Dirac mass matrix for the type-I seesaw.

In the bidoublet sector, the SM Higgs $h$ is predominantly from the real component of the neutral scalar $\phi_1^0$. There is a heavy CP-even scalar $H_1$ from the real component of $\phi_2^0$, which couples to the SM quarks through the couplings
\begin{eqnarray}
\label{eqn:Lyukawa}
-{\cal L}_Y \ & \supset & \
h_{q}\overline{Q}_{L}\Phi Q_{R}
+ \tilde{h}_{q} \overline{Q}_{L} \widetilde{\Phi} Q_{R} \,,
\end{eqnarray}
with $q_{L,R} = (u,\,d)^{\sf T}_{L,R}$ the left- and right-handed quark doublets, $\widetilde{\Phi}=i\sigma_2\Phi^*$ (with $\sigma_2$ being the second Pauli matrix), and $h_q$ and $\tilde{h}_q$ the quark coupling matrices. After symmetry breaking, the tree-level couplings of $H_1$ to the SM quarks are flavor-changing, which are governed by the quark masses and the left- and right-handed quark mixing matrices $V_{L,\,R}$ in the form of
\begin{eqnarray}
-{\cal L}_Y \supset
 H_1^0 \bar{d}_i d_j \left[
 - \sqrt2 \xi \widehat{Y}_D
   + \frac{1}{\sqrt2} \left( V_L^\dagger \widehat{Y}_U V_R \right) \right]_{ij} \,,
\end{eqnarray}
with $\xi = \kappa'/\kappa$ the VEV ratio in the bidoublet sector, $i,\,j$ the quark generation indices, and $\widehat{Y}_{U,D}$ diagonal Yukawa coupling matrices for the SM up- and down-type quarks. The tree-level FCNC couplings of $H_1$ contribute significantly to the neutral $K$ and $B$ meson oscillations, and thus $H_1$ is required to be superheavy, roughly above 15 TeV~\cite{Zhang:2007da, Bertolini:2014sua, Bertolini:2019out}.

The CP-even neutral component $S$ from the triplet $\Delta_R$ couples predominantly to the BSM scalars, heavy $W_R$ and $Z_R$ gauge bosons and the heavy RHNs in the LRSM, and all the couplings of $S$ to the SM particles are from its mixings with the SM Higgs $h$ and heavy $H_1$~\cite{Zhang:2007da, Dev:2016dja}. Therefore in some region of the parameter space, even if the radiative corrections to $S$ mass are taken into consideration, $S$ can be very light, e.g. in the sub-GeV-scale~\cite{Dev:2016vle, Dev:2017dui}. Thus it might be a good candidate to explain the KOTO anomaly.

\subsection{Fitting the KOTO anomaly} \label{sec:koto2}

In the LRSM, the FCNC couplings of $S$ are from mixing with the SM Higgs $h$ and the heavy scalar $H_1$ from the bidoublet.  Denoting these mixing angles  respectively by $\sin\theta_{1,\,2}$, the FCNC couplings of $S$ to $s$ and $d$ quarks will be proportional to the factor of $ \left( \xi\sin\theta_1 + \sin\theta_2 \right) \big( V_L^\dagger \widehat{Y}_U V_R \big)_{12}$, {where the right-handed quark mixing matrix $V_R$ is almost the same as the CKM matrix $V_L$ in the SM, up to some ambiguous signs~\cite{Zhang:2007fn, Zhang:2007da}. For the sake of concreteness we set explicitly $V_R = V_L$ throughout this paper.}
Note that as a result of the $CP$ phase in the $V_{L,\,R}$ matrices, this coupling is complex. The partial widths for the charged and neutral $K$ meson decays are given by~\cite{Dolan:2014ska,Izaguirre:2016dfi,He:2006uu}
\begin{widetext}
\begin{eqnarray}
\label{eqn:Kdecay1}
\Gamma (K^\pm \to \pi^\pm S) & \ = \ &
\frac{G_F m_{K^\pm} \left( \xi\sin\theta_1 + \sin\theta_2 \right)^2}{8\sqrt2 \pi}
\frac{m_{K^\pm}^2}{m_S^2}
\left|  \left( V_R^\dagger \widehat{M}_U V_L \right)_{21} \right|^2
\left( 1 - \frac{m_{\pi^\pm}^2}{m_{K^\pm}^2} \right)^2
\beta_2 (m_{K^\pm}, m_{\pi^\pm}, m_{S}) \,, \\
\label{eqn:Kdecay2}
\Gamma (K_L \to \pi^0 S) & \ = \ &
\frac{G_F m_{K_L} \left( \xi\sin\theta_1 + \sin\theta_2 \right)^2}{8\sqrt2 \pi}
\frac{m_{K^0}^2}{m_S^2}
\left|  {\rm Re}\left( V_R^\dagger \widehat{M}_U V_L \right)_{21} \right|^2
      \left( 1 - \frac{m_{\pi^0}^2}{m_{K_L}^2} \right)^2
\beta_2 (m_{K_L}, m_{\pi^0}, m_{S}) \,,
\end{eqnarray}
\end{widetext}
with the kinematic function $\beta_2(M,m_1,m_2)$ defined in Eq.~\eqref{eqn:beta2}.

The mixing angles of $S$ to $h$ and $H_1$ are strongly constrained by the low-energy high-precision flavor data, depending on the $S$ mass in the LRSM~\cite{Dev:2017dui}. At the one-loop level, $S$ can decay into two photons, i.e. $S \to \gamma\gamma$, which is induced by the $W_R$ boson and the singly- and doubly-charged scalars~\cite{Dev:2016vle, Dev:2017dui}:
\begin{eqnarray}
\label{eqn:H3diphoton}
\Gamma (S \to \gamma\gamma) & \ \simeq \ &
\frac{\alpha^2 m_{S}^3}{18 \pi^3 v_R^2} \,,
\end{eqnarray}
where we have neglected the contributions from the SM fermion loops which are all highly suppressed by the mixing angles $\sin\theta_{1,2}$, and take the limit of light $S$ (compared to the BSM particles in the loop). Note that the partial width does not depend on the gauge coupling $g_R$, as the dependence of $W_R$ couplings and $W_R$ mass on $g_R$ are canceled out. Thus, the partial width of $S$ to diphoton is effectively suppressed only by the $v_R$ scale, independent of the mixing angles $\theta_{1,2}$.

\begin{figure*}[!t]
  \centering
  \includegraphics[width=0.65\textwidth]{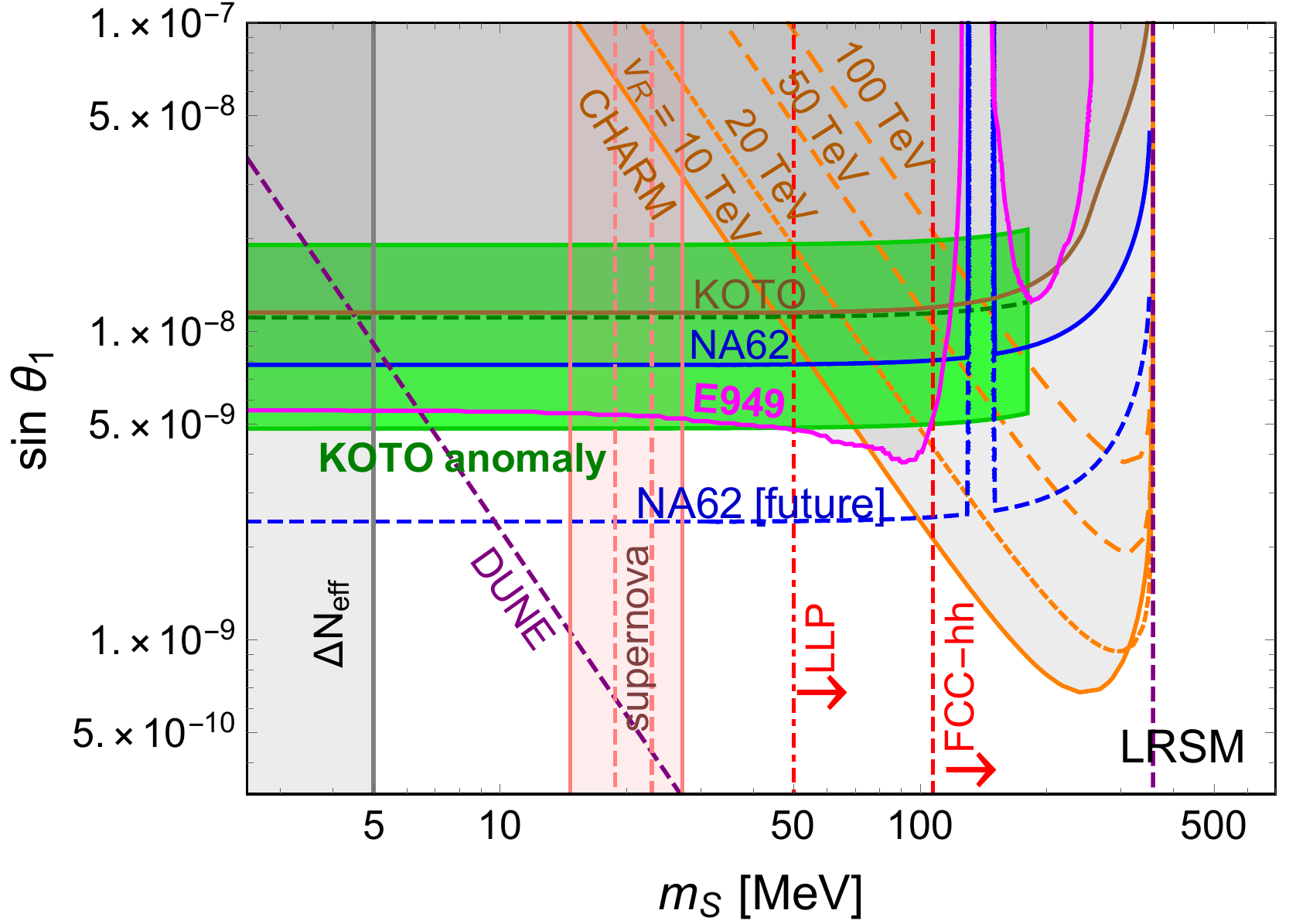}
  \caption{The parameter space favored by the KOTO anomaly~\cite{anomaly} in the LRSM (cf.~Sec.~\ref{sec:koto2}) is shown by the green shaded region (95\% CL), with the green dashed line corresponding to the central value quoted in Eq.~\eqref{eqn:anomaly}. For comparison, we show the exclusion regions from a previous KOTO search for $K_L\to \pi^0\nu\bar{\nu}$ (brown)~\cite{Ahn:2018mvc}, NA62 search for $K^+ \to \pi^+ \nu \bar\nu$ (blue)~\cite{NA62}, E949 search for $K^+ \to \pi^+ X$ (magenta)~\cite{Artamonov:2009sz}, and beam-dump experiment at CHARM (orange)~\cite{Bergsma:1985qz}; cf.~Sec.~\ref{sec:lab2}. All the gray shaded regions are excluded. The grey and pink shaded vertical regions are excluded by the $\Delta N_{\rm eff}$ and supernova constraints respectively (cf.~Sec.~\ref{sec:ac2}). Also shown are the future prospects at NA62 (dashed blue)~\cite{Anelli:2005ju}, DUNE (dashed purple)~\cite{Adams:2013qkq}, the long-lived particle searches at LHC (dashed red) and FCC-hh (dot-dashed red)~\cite{Dev:2017dui}.
 For the CHARM limits we choose four benchmark values of $v_R = 10$, 20, 50 and 100 TeV. Similarly, the solid and dashed lines for the supernova limits correspond respectively to the luminosity of $2 \times 10^{53}$ erg and $3 \times 10^{53}$ erg for a fixed $v_R=10$ TeV (cf. Fig.~\ref{fig:supernova}).}
  \label{fig:LRSM}
\end{figure*}

As detailed in Refs.~\cite{Dev:2016vle, Dev:2017dui}, if $S$ is below the GeV-scale, it tends to be long-lived. In the limit of $\sin\theta_{1,\,2} \to 0$, the dominant decay mode of $S$ is the diphoton channel, and its lifetime only depends on its mass $m_{S}$ and the $v_R$ scale [cf.~Eq.~\eqref{eqn:H3diphoton}]. A long-lived $S$ in the LRSM with lifetime $bc\tau_S \gtrsim 3$ m can be  a good candidate for the KOTO anomaly. Setting $\sin\theta_{2} = 0$, the preferred parameter space of $m_{S}$ and the $S-h$ mixing angle $\sin\theta_{1}$ for the KOTO anomaly is shown by the shaded green region in Fig.~\ref{fig:LRSM}.\footnote{The other choice, namely, setting  $\sin\theta_{1} = 0$ yields a very similar plot in the $(m_S,\sin\theta_2)$ plane, with the mixing angle $\sin\theta_2$ smaller than $\sin\theta_1$ in Fig.~\ref{fig:LRSM} by a factor of $\kappa'/\kappa = m_b/m_t$~\cite{Dev:2016vle, Dev:2017dui}, and is therefore not shown here.} As in Fig.~\ref{fig:U1}, the dashed green line corresponds to the central value of the KOTO result, while the shaded green band is the 95\% CL favored region from Eq.~(\ref{eqn:anomaly}).  The region with $S$ mass $m_{S} > 180$ MeV is not included here, because it does not have any overlap with the KOTO signal region~\cite{Kitahara:2019lws}. For concreteness, we set the VEV ratio $\xi = \kappa'/\kappa = m_b/m_t$ which is natural for the known hierarchy of bottom and top quark masses. We have evaluated the effective branching ratio in Eq.~(\ref{eqn:BReff}) for different $v_R$ values and found that it is almost independent of the $v_R$ value, as in the parameter space of interest the typical lifetime of $S$ is much longer than the KOTO detector size of 3 m. As the FCNC couplings of $S$ are at tree-level in the LRSM, the mixing angles $\sin\theta_{1}$ for the KOTO anomaly (and the following constraints) are orders of magnitude smaller than in the generic singlet model (cf. Fig.~\ref{fig:U1}).

\subsection{Laboratory constraints} \label{sec:lab2}

As for the generic singlet model in Section~\ref{sec:generic}, the most stringent limits for the parameter space relevant for the KOTO anomaly are from the searches of $K^+ \to \pi^+ \nu \bar\nu$ at  NA62~\cite{NA62}, $K^+ \to \pi^+ X$ at E949~\cite{Artamonov:2009sz}, $K_L \to \pi^0 X$ at KOTO~\cite{Ahn:2018mvc}, and the $e^+ e^-$, $\mu^+ \mu^-$ and $\gamma\gamma$ decay products from kaon decay at the CHARM beam-dump experiment~\cite{Bergsma:1985qz}. Evaluations of these limits are quite similar to those in the generic singlet model, as discussed in Sec.~\ref{sec:lab1} and we do not repeat them here. As in Fig.~\ref{fig:U1}, the current E949, NA62 and KOTO limits are shown respectively by the magenta, blue and brown lines in Fig.~\ref{fig:LRSM}, and the future NA62 improvement is indicated by the dashed blue lines, which corresponds to a limit down to $2.35 \times 10^{-11}$ for the branching ratio ${\rm BR} (K^+ \to \pi^+ S)$~\cite{Anelli:2005ju}.

The limits from the CHARM beam-dump experiment are presented by the orange line in Fig.~\ref{fig:LRSM}. Unlike the generic singlet model, the most stringent CHARM limit in the LRSM comes from the $\gamma\gamma$ channel, since this is the dominant decay mode of $S$ for small mixing angles. Therefore the event number depends on the $v_R$ scale, as illustrated with four benchmark values of $v_R = 10$, 20, 50 and 100 TeV. For a larger $v_R$ value, the lifetime of $S$ tends to be longer and as a result the CHARM limits get weaker. With an improved proton-on-target number (PoT) of $5\times 10^{21}$, DUNE can collect $8 \times 10^{21}$ kaons, with $M_{pp} = 11$ and $\chi_s = 1/7$~\cite{Adams:2013qkq}. With the energy $E_{S} \simeq 12$ GeV, the decay length parameters $L = 500$ m and $\Delta L = 7$ m for the DUNE beam dump set up~\cite{Adams:2013qkq}, and setting the Kaon absorption length at DUNE the same as that for CHARM, the current CHARM limits on the mixing angle $\sin\theta_{}$ can be improved by two orders of magnitude, as shown by the dashed purple curve in Fig.~\ref{fig:LRSM}. The decay $K \to \pi S$ can also be searched for in the SHiP experiment, but the PoT number $2\times 10^{20}$ is almost one order of magnitude lower than DUNE, and the lifetime that can be probed is also shorter~\cite{Alekhin:2015byh}. Thus, we estimate that the prospect of $S$ search at SHiP is weaker than at CHARM and DUNE~\cite{Dev:2017dui} and is not shown in Fig.~\ref{fig:LRSM}.


For all the calculations above in the LRSM, we have set the VEV ratio $\xi = \kappa'/\kappa = m_b/ m_t$. When the $\xi$ parameter is different, the KOTO region, the NA62 and CHARM limits for $\sin\theta_1$ are all universally rescaled by $\xi$, and this does not help to enlarge the parameter space for the KOTO anomaly. As for the generic singlet model, the limits from the flavor-changing decays $K \to \pi \chi\chi$ (with $\chi\chi = e^+ e^-,\, \mu^+ \mu^- ,\, \gamma\gamma$) are not applicable to the long-lived $S$, and the limits from $B$ meson decays, $K$ and $B$ meson oscillations are much weaker than those from the $K$ mesons in the parameter space of interest.

As can be seen from Eqs.~(\ref{eqn:Kdecay1}) and (\ref{eqn:Kdecay2}), the decay $K \to \pi + S$ for the KOTO anomaly and the KOTO, E949 and NA62 limits are determined by the scalar mass $m_S$ and the mixing angle $\sin\theta_1$, whereas the CHARM limit are mostly from the decay $S \to \gamma\gamma$ which is dictated by the scalar mass $m_S$ and the $v_R$ scale in the limit of small mixing angles [cf.~Eq.~(\ref{eqn:H3diphoton})]. Therefore the LRSM could accommodate the KOTO anomaly while evading the stringent limits from CHARM (and the supernova limits below), which is very different from the singlet scalar model in Section~\ref{sec:generic}.

As in the generic singlet case in Section~\ref{sec:generic}, the long-lived scalar $S$ in the LRSM can also be searched for as LLP in the high-energy colliders~\cite{Dev:2017dui}. Unlike  the singlet scalar case, when  the mixing angles $\sin\theta_{1,2}$ are very small (cf. Fig.~\ref{fig:LRSM}), the scalar $S$ in the LRSM can be produced from the gauge interactions mediated by the heavy $W_R$ (and $Z_R$) bosons, i.e. $pp \to W_R (Z_R)S$.
As a result of the Majorana nature of the heavy RHNs in the LRSM, the smoking-gun signal of $W_R$ boson at hadron colliders is same-sign dilepton plus jets without significant missing energy~\cite{Keung:1983uu}, and the current most stringent LHC same-sign dilepton limits requires that the $W_R$ mass $m_{W_R} \gtrsim 5$ TeV, depending on the RHN mass~\cite{Aaboud:2019wfg}.\footnote{Even if the RHNs are heavier than the $W_R$ boson, there are also the direct LHC searches of $W_R \to t\bar{b}$, which exclude $W_R$ mass below 3.25 TeV~\cite{Aaboud:2018jux}.} For a 5 TeV $W_R$ boson, the production cross section of $S$ at LHC 14 TeV is only $\sigma (pp \to W_RS) \simeq 0.025$ fb, and we cannot have any LLP prospects for $m_S < 1$ GeV at LHC or MATHUSLA if the  $SU(2)_R$ gauge coupling is the same as that for $SU(2)_L$ (see the left panel of Fig.~17 in Ref.~\cite{Dev:2017dui}). It is easy to understand: in the limit of small mixing angles $\sin\theta_{1,2}$, the decay width is proportional to $m_S^3/v_R^2$ (cf. Eq.~(\ref{eqn:H3diphoton})); so for a light $S$, the decay lifetime is so long that almost no $S$ decays inside the LHC detector. At future 100 TeV colliders FCC-hh and SPPC, the production cross section of $S$ can be almost four orders of magnitude larger than at LHC 14 TeV for $m_{W_R} = 5$ TeV, and we can have LLP prospects for below-GeV scale at FCC-hh and the dedicated LLP detectors therein~\cite{Dev:2017dui}. Setting $m_{W_R} = 5$ TeV, the LLP prospects at FCC-hh and the dedicated LLP detector is shown in Fig.~\ref{fig:LRSM} respectively by the dashed and dot-dashed red lines. The regions to the right side of the red lines can be probed by the LLP searches, which however do not have any allowed KOTO signal region. 

\subsection{Astrophysical and cosmological constraints} \label{sec:ac2}

\begin{figure}[!t]
  \centering
  \includegraphics[width=0.48\textwidth]{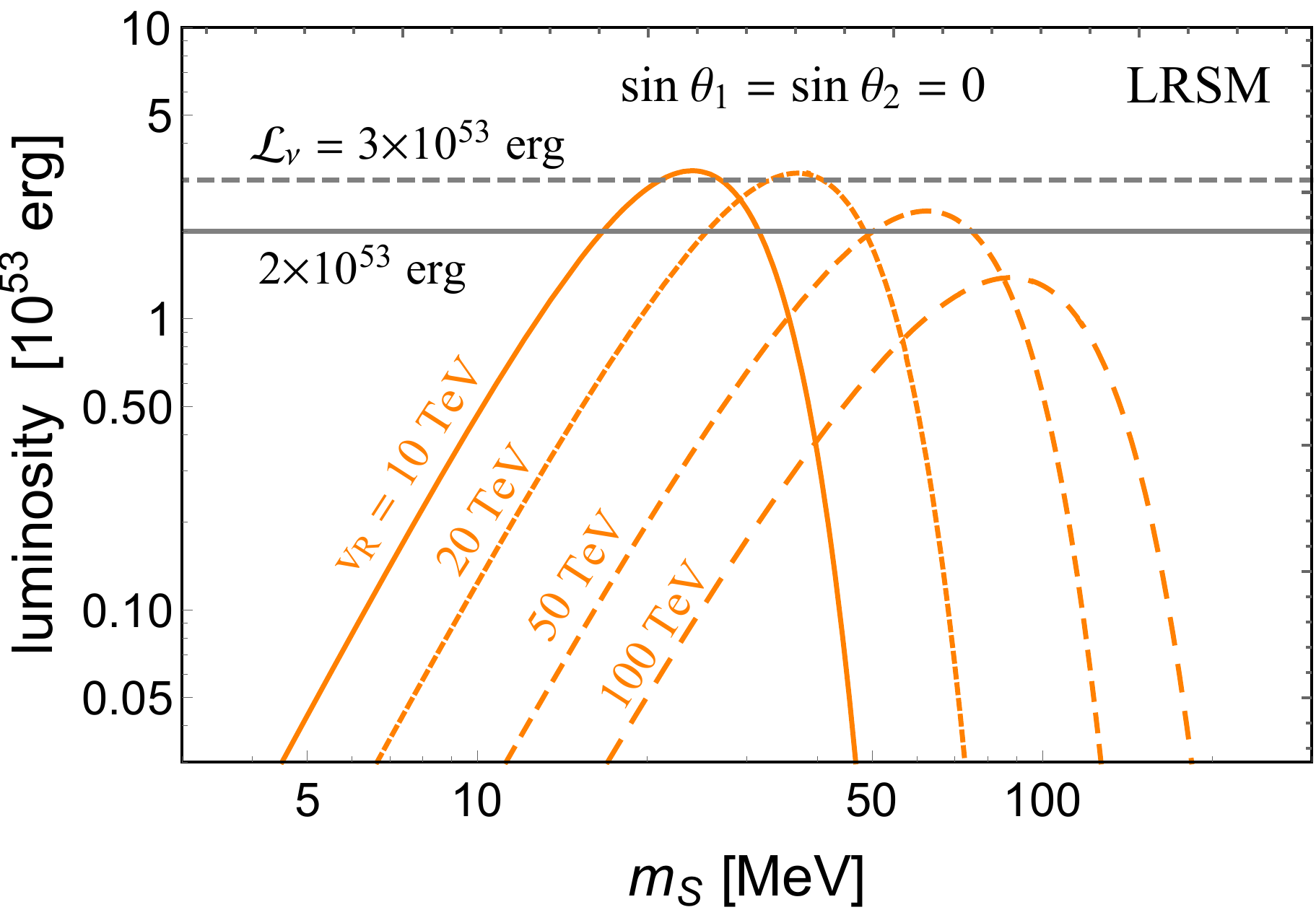}
  \caption{Supernova limits on the light scalar $S$ in the LRSM as function of its mass for different values of $v_R=10$, 20, 50 and 100 TeV. The horizontal solid and dashed grey lines indicate respectively the total energy loss of $2\times 10^{53}$ erg and $3\times 10^{53}$ erg due to neutrino emission. Here we have taken both the $S$ mixing angles $\theta_1$ (with the SM Higgs) and $\theta_2$ (with the heavy bidoublet) to be zero, so only the gauge interactions are relevant. }
  \label{fig:supernova}
\end{figure}

As in the generic singlet model case, if $S$ is light, it can be produced in the supernova core and get constrained by the collapse luminosity. In the LRSM, $S$ can be produced in two distinct channels:
\begin{enumerate}[(i)]
  \item Nuclear bremsstrahlung process ${\cal N} + {\cal N} \ \to \ {\cal N} + {\cal N} + S$, which originates from the mixing with the SM Higgs. In this case, the effective couplings of $S$ to nucleons are highly suppressed by the mixing angle $\sin\theta_1$ required for the KOTO explanation, thus the corresponding supernova limits are too weak and not relevant to the KOTO anomaly. For the
  same reason, the reabsorption contribution to mean free path of the $S$ is also suppressed.

  \item Photon fusion process, i.e. $\gamma\gamma\to S$, which is highly suppressed by the ratio $m_{S}^2/v_R^2$ [cf. Eq.~(\ref{eqn:H3diphoton})]. Assuming the photon momentum follows the Bose-Einstein distribution in  the supernova core, we follow the calculations in Ref.~\cite{Heurtier:2016otg} to estimate the production rate of $S$ which turns out to be just at the order of $\sim 10^{53}$ erg for the benchmark values of $v_R = (10 - 100)$ TeV, as shown in Fig.~\ref{fig:supernova}. For simplicity, we have set both the scalar mixing angles $\sin\theta_{1,2}$ to be zero. The region of $m_S$ for which the luminosity exceeds the observed value of $(2-3)\times 10^{53}$ erg~\cite{Hirata:1987hu} can be excluded. For instance, the supernova limits for $v_R = 10$ TeV are shown by the pink shaded region in Fig.~\ref{fig:LRSM}, with the solid and dashed lines corresponding to the luminosity of $2 \times 10^{53}$ erg and $3 \times 10^{53}$ erg respectively, which exclude respectively the mass ranges of $15 \, {\rm MeV} \lesssim m_S \lesssim 27$ MeV and $19 \, {\rm MeV} \lesssim m_S \lesssim 23$ MeV. If $v_R$ goes higher than roughly $\sim 50$ TeV, the production rate will be too small such that we do not have any supernova limits.
\end{enumerate}

In the LRSM, even if the mixing angles $\sin\theta_{1,2}$ are extremely small, $S$ can still decay into two photons through the $W_R$ boson and the heavy charged scalars. Therefore in the parameter space of interest, the lifetime of $S$ is always much shorter than one second, and we do not have any limits from BBN.

As in the $U(1)_{B-L}$ model case, a light $S$ contributes to the relativistic degree of freedom $N_{\rm eff}$ in the early universe. In the limit of small mixing angles $\sin\theta_{1,\,2}$, $S$ can be in equilibrium with photon if its mass $m_\phi \gtrsim 2$ MeV~\cite{Dev:2017dui}. The current limit of $\Delta N_{\rm eff} < 0.7$ has excluded a scalar particle lighter than 5 MeV~\cite{Nollett:2014lwa, Kamada:2015era}, which is shown by the gray shaded region in Fig.~\ref{fig:LRSM}.

As can be seen from Fig.~\ref{fig:LRSM}, although the central value of the KOTO anomaly has been excluded mostly by E949 and NA62, there is still a narrow band left within the $2\sigma$ uncertainty of KOTO data between $5~{\rm MeV}\lesssim m_S\lesssim 46~{\rm MeV}$ with $\sin\theta_1\sim (5-6)\times 10^{-8}$, even after all the laboratory and cosmological/astrophysical limits are taken into consideration. In addition, the full parameter space can be conclusively tested in the future NA62 and DUNE data.

\section{Conclusion} \label{sec:con}

The three tantalizing events found in the signal region of the flavor-violating decay $K_L \to \pi^0 \nu \bar\nu$ at the KOTO experiment  might be a glimpse of BSM physics. Possible explanation of this by a light long-lived scalar particle which has either tree or loop-level flavor-changing couplings to the $s$ and $d$ quarks and has a lifetime approximately larger than the KOTO detector size of 3 m has been suggested~\cite{Kitahara:2019lws}. In this paper, we have studied three possible model implications of this suggestion and constraints on the model parameters from various laboratory measurements and astrophysical/cosmological observations to see if there is any parameter space left to explain this anomaly.
In the SM+$S$ model and $U(1)_{B-L}$ model, there is a narrow range of parameters for {$110 \, {\rm MeV} \lesssim m_S \lesssim 180$ MeV and $2.5 \times 10^{-4} \lesssim {\rm sin}^\theta \lesssim 6.5 \times 10^{-4}$} which satisfies all the laboratory constraints  and where the KOTO anomaly can be explained, {as seen in Fig.~\ref{fig:U1}}.
Similarly, in  the LRSM,  there also remains a narrow range of parameter space between $5~{\rm MeV}\lesssim m_S\lesssim 46~{\rm MeV}$ with Higgs mixing angle $\sin\theta_1\sim (5-6)\times 10^{-8}$ which can explain the KOTO anomaly within the $2\sigma$ range, while being consistent with all existing constraints, as shown in Fig.~\ref{fig:LRSM}. In both these models, this allowed parameter space can be tested in future NA62 and DUNE experiments.

\vspace{0.2cm}
\noindent
{\bf Note Added:} While finalizing our manuscript, we noticed Ref.~\cite{Egana-Ugrinovic:2019wzj}, which has some overlap with our SM-singlet scalar case (cf.~Sec.~\ref{sec:generic}).

\section*{Acknowledgments}

We would like to thank Jim Cline for pointing out a typo in Eq.~(\ref{eqn:ysd}) of the earlier version. We are very grateful to Daniel Egana-Ugrinovic, Sam Miller and Patrick Meade for useful correspondence and comparison with the results in Ref.~\cite{Egana-Ugrinovic:2019wzj}, which helped us correct our earlier calculations of the flavor-changing Kaon decays for the singlet scalar case, including the effective FCNC scalar couplings, the threshold for the hadron decays and the re-absorption factor, {as well as the impact of the re-absorption process on the supernova limits}. We also thank Kaladi Babu, Sabyasachi Chakraborty, Bhaskar Dutta and Taku Yamanaka for useful discussions and correspondence. Y.Z.  thanks the High Energy Theory  Group  at  Oklahoma  State  University  for  warm hospitality, where part of this work was done. Y.Z. is also grateful to the Center for Future High Energy Physics, Institute of High Energy Physics, Chinese Academy of Sciences, where the work was updated. The work of B.D. and Y.Z. is supported in part by the US Department of Energy under Grant No.  DE-SC0017987 and in part by the MCSS. This work was also supported in part by the Neutrino Theory Network Program under Grant No.  DE-AC02-07CH11359. The work of R.N.M. is supported by the US National Science Foundation grant no. PHY-1914631.

\end{document}